\documentstyle[aps,prl]{revtex}
\begin{document}
\draft
\twocolumn[\hsize\textwidth\columnwidth\hsize\csname %
@twocolumnfalse\endcsname

\title {Electron Green's function in the planar $t-J$ model}
\author{P. Prelov\v sek }
\address{ J. Stefan Institute, University of Ljubljana, 1001
Ljubljana, Slovenia }
\date{\today}
\maketitle
\begin{abstract}
\widetext
{\it Dedicated to Prof. Wolfgang G\"otze on the occasion of his}
$60^{th}$ {\it birthday}.

\smallskip
The electron Green's functions $G({\bf k},\omega)$ within the $t-J$
model and in the regime of intermediate doping is studied analytically
using equations of motion for projected fermionic operators and the
decoupling of the self energy into the single-particle and spin
fluctuations. It is shown that the assumption of marginal spin
dynamics at $T=0$ leads to an anomalous quasiparticle damping.
Numerical result show also a pronounced asymmetry between the hole
($\omega<0$) and the electron ($\omega>0$) part of the spectral
function, whereby hole-like quasiparticles are generally overdamped.

\end{abstract}
]
\narrowtext
\section{Introduction}

The understanding of low-energy excitations in metals with strongly
correlated electrons is at present one of the central challenges
within the solid-state physics. Theoretical investigations have been
to large extent motivated by the observation of anomalous electronic
properties of superconducting cuprates.  We shall discuss here only
the state of the `strange' metal, as manifested in cuprates at
$T>T_c$.  Our study is devoted to the single-particle response, which
is in cuprates investigated directly by the angle resolved
photoemission (ARPES) probing the spectral function $A({\bf k},
\omega)$ \cite{shen}. ARPES reveals in compounds with intermediate
doping a quasiparticle (QP) dispersion consistent with a large
electronic Fermi surface. On the other hand the conclusions on the QP
character, as deduced from the low-energy $\omega \sim 0$ spectral
properties, are less clear. Still it seems that spectral shapes are
never underdamped, as expected for the QP excitations near the Fermi
surface within the usual Fermi liquid (FL) theory.  In order to
explain the anomalous response, a phenomenological marginal Fermi
liquid (MFL) theory has been proposed \cite{varm,litt}, which assumes
at low $T$ a frequency-dependent QP damping of the form
$\Sigma''(\omega) \propto \omega$, rather than the FL behaviour
$\Sigma''(\omega) \propto \omega^2$.  While for ARPES the
interpretation of results is still controversial
\cite{shen}, there is more agreement on the evidence for the anomalous
spin dynamics, as manifested within the NMR relaxation experiments
\cite{imai} and the neutron scattering\cite{shir}, and for the anomalous
charge dynamics, as deduced from the resistivity $\rho(T) \propto T$
and from the non-Drude form of the optical conductivity \cite{rome}.
Mentioned anomalies seem to be well accounted by the MFL scenario
\cite{varm,litt}. 

Theoretical investigations of electron Green's functions and related
spectral properties, starting on the level of prototype microscopic
models for correlated electrons, as the Hubbard model or the $t-J$
model, have reached quite a level of consensus for a single mobile
carrier (hole) introduced into the reference insulator \cite{brin},
representing the spin polaron in the case of the antiferromagnetic
(AFM) spin background \cite{schm}.  Even here there are open questions
in the relation of model results with relevant ARPES experiments
\cite{well}. While single-hole results could be possibly extended to
the low-doping regime, the spectral properties of the
intermediate-doping regime are much more difficult to approach
theoretically. So far there has been a number of numerical studies,
where $A({\bf k}, \omega)$ has been calculated via the exact
diagonalization and the quantum Monte Carlo methods in small systems
\cite{step,bulu,dago}. Whereas these investigations obtained
valuable information on the QP dispersion and on the location of the
Fermi surface (FS), due to several restrictions it was not possible to
resolve spectral widths and shapes, so results were quite
restricted in determining the low-energy QP properties, being at the
core of different scenarios for `strange' metal.  Recently a novel
numerical method \cite{jakl1}, combining Lanczos diagonalization and
random sampling for finite systems at $T>0$, has been applied to the
study of $A({\bf k}, \omega)$ within the $t-J$ model \cite{jakl2}. It
was possible to extract self energies $\Sigma({\bf k}, \omega)$, which
revealed the anomalous behavior following the MFL scenario, i.e. for
$k \sim k_F$ it was found $\Sigma''({\bf k}, \omega) \propto |\omega|
+ \xi T $. This finding is consistent with the MFL-type charge and
spin dynamics, as found previously in numerical studies within the
$t-J$ model \cite{jakl3}, as well as in experiments on cuprates
\cite{imai,rome}.

There have been few attempts to treat electron Green's functions
analytically at intermediate doping. Starting from the one-band
Hubbard model, self energies have been related to the AFM spin
fluctuations within the random phase approximation and weak
coupling \cite{kamp}, and within a selfconsistent (FLEX) theory
\cite{bick}.  Both approaches rely on the assumption of modest
correlations, i.e. not too large $U/t$.  At least within the weak
coupling the analysis yields a low-energy behaviour according to
normal FL. Spectral functions in the strong-correlation regime, where
the $t-J$ model should be more appropriate starting point, have proven
to be even harder due to projections involved in fermionic operators
and due to composite character of electrons within the slave-boson
theories \cite{wang}.

In this paper we introduce a simple theory for the electron Green's
function $G({\bf k},\omega)$ in the strong-correlation regime.  The
central observation is that within the $t-J$ model one could work
directly with projected fermionic operators. Since these prevent the
application of usual diagrammatic techniques, we apply the method of
equations of motion to express the self energy $\Sigma({\bf
k},\omega)$. Working at $T=0$ we divide $\Sigma$ into a contribution
coherent at $\omega \to 0$, and an incoherent part. We approximate the
former by performing a decoupling into the single-particle propagation
and spin fluctuations.  We do not intend to evaluate within the same
framework the spin dynamics, so we assume it of the MFL-type form, as
found in previous studies \cite{jakl2}. In spite of its simplicity such
an analysis yields promising features.

\section{Projected electron propagator}

In the following we study the $t-J$ model \cite{rice} as a prototype
model for strongly correlated electrons, and for electronic properties
of cuprates in particular,
\begin{equation}
H=-t\sum_{\langle ij\rangle s}(\tilde{c}^\dagger_{js}\tilde{c}_{is}+
\text{H.c.})+J\sum_{\langle ij\rangle}{\bf S}_i\cdot {\bf S}_j ,
\label{eq1}
\end{equation}
where we have omitted usual but less important density-density
coupling in the $J$ term, and
\begin{equation}
\tilde{c}^\dagger_{is}= (1-n_{i,-s}) c^\dagger_{is} \label{eq2}
\end{equation}
are local fermionic operators, which project out the states with the
double occupancy. ${\bf S}_i={1\over 2}\sum_{s s'} c^\dagger_{is}
\hbox{\boldmath$\sigma$}_{ss'} c_{is'} $ are local spin operators.

Our aim is to calculate the electron propagator
\begin{eqnarray}
G({\bf k},z)&=& \langle\!\langle \tilde c_{{\bf k} s}; \tilde
c^{\dagger}_{{\bf k} s} \rangle\!\rangle _{z} = \nonumber \\
&=&-i \int_0^{\infty}
{\mathrm e}^{i(z+\mu) t}\langle \{ \tilde c_{{\bf k} s}(t) ,
\tilde c^{\dagger}_{{\bf k} s} \}_+  \rangle dt, \label{eq3}
\end{eqnarray}
where $z=\omega + i\delta,~\delta > 0$, $\mu$ is the chemical
potential, and
\begin{equation}
\tilde c^{\dagger}_{{\bf k} s} = {1\over \sqrt{N}} \sum_i 
{\mathrm e}^{i{\bf k} 
\cdot {\bf r}_i} \tilde c^{\dagger}_{i s}.  \label{eq4}
\end{equation}

Model Eq.(\ref{eq1}) is defined within the basis set, which
does not contain doubly occupied sites. Within this subspace the
Green's function, Eq.(\ref{eq2}), expressed with projected operators,
is the same as the usual electronic propagator.  Since commutation
relations for projected operators Eq.(\ref{eq2}) are not simple, the usual
diagrammatic derivation of the fermionic self energy is not
applicable.  Hence we employ the method of equations of motion,
introduced to treat dynamics of more general operators \cite{zuba}.

Note that general (anticommutator) correlation functions obey the
equations of motion \cite{zuba}
\begin{eqnarray}
z \langle \!\langle A;B \rangle \!\rangle_z &=& \langle
 \{A,B\}_+\rangle + \langle \!\langle [A,H]; B \rangle \!\rangle_z =
 \nonumber \\ &=& \langle \{A,B\}_+\rangle - \langle \!\langle A;[B,H]
 \rangle \!\rangle_z.
\label{eq5}
\end{eqnarray}
Let us apply Eqs.(\ref{eq5}) to the autocorrelation function $G(z) =
\langle\!\langle A;A^+ \rangle\!\rangle_z $. If we define the 
operator $C$ as
\begin{equation}
[A,H]= \zeta A - i C, \qquad \langle \{C, A^+\}_+\rangle =0, \label{eq6}
\end{equation}
we can express $G$ via Eq.(\ref{eq5}),
\begin{eqnarray}
G(z) &=& G_0(z) + {1\over \alpha^2} G_0(z)^2 \langle\!\langle C;C^+
\rangle\!\rangle_z , \nonumber \\ G_0(z)&=& {\alpha \over z-\zeta}, 
\qquad
\alpha=\langle \{A,A^+\}_+\rangle. \label{eq7}
\end{eqnarray}
It is an evident intention to reexpress Eq.(\ref{eq7}) in terms of 
self energy $\Sigma(z)$
\begin{equation}
G(z)={\alpha\over z - \zeta -\Sigma(z) },\qquad \Sigma(z) \sim {1\over
 \alpha} \langle\!\langle C;C^+ \rangle\!\rangle_z^{irr}. \label{eq8}
\end{equation}
Such an expression for $\Sigma(z)$ would within the diagrammatic
techniques correspond to the contribution of 'irreducible' diagrams.
It is valid within the perturbation theory, as applied e.g. to the
analysis of the dynamical conductivity \cite{gotz}. More generally it
could be considered as a memory function in analogy with the Mori's
projection method \cite{mori}, where again only the `irreducible' part
of $\langle\!\langle C;C^+ \rangle\!\rangle_z^{irr}$ would contribute
to $\Sigma(z)$.  Recently a treatment of Green's functions, analogous
to Eqs.(\ref{eq5}-\ref{eq8}), has been applied also to some problems
of correlated electrons \cite{plak1,plak}.

The `irreducible' contribution is in the present analysis approximated
by an appropriate decoupling on this level, in the sense of
mode-coupling theories. It is evident that in our application the
decoupling is justified in terms of underlying physics, since the
theory does not have a small parameter.

The application of Eqs.(\ref{eq3}-\ref{eq7}) to the electron $G({\bf
k},\omega)$ is straightforward. We define in analogy to 
Eq.(\ref{eq8}),
\begin{equation}
G({\bf k},z)= {\alpha_{\bf k} \over z +\mu -\zeta_{\bf k} -
\Sigma({\bf k},z) },\label{eq9} 
\end{equation}
and the corresponding spectral function $A({\bf k},\omega) = 
-(1/\pi){\rm Im} G({\bf k},\omega)$. First we note that
\begin{eqnarray}
\alpha_{\bf k} &=& \langle \{\tilde c_{{\bf k} s},\tilde 
c^{\dagger}_{{\bf k} s}\}_+ \rangle =
 {1\over N}\sum_i \langle \{\tilde c_{i s},\tilde c^{\dagger}_{i
s}\}_+ \rangle = \nonumber \\
 &=& 1 - {c_e \over 2} = {1\over 2} (1+c_h),
\label{eq10}
\end{eqnarray}
where $c_e, c_h=1-c_e$ are the electron and the hole concentration,
respectively. Eqs.(\ref{eq8},\ref{eq10}) imply that within the $t-J$
model the spectral function is not normalized to unity
\cite{step,jakl2}, as the consequence of the projected fermionic
basis. Still the normalization constant $\alpha$ is $\bf
k$-independent. It should be also noted that due to $\alpha <1$ the
self energy $\Sigma$ in Eq.(\ref{eq9}) does not coincide with the
standard definition $\bar \Sigma$ which requires $\alpha=1$ in the
numerator of Eq.(\ref{eq9}), but can be related to it, i.e. $\bar
\Sigma = [\Sigma + (\alpha -1)z]/\alpha$. Note however that $\bar
\Sigma$ would be less convenient since $\bar \Sigma(\omega \to 
\pm \infty) \ne 0$ \cite{jakl2}.

Next we consider the equations of motion
\begin{eqnarray}
[\tilde c_{is},H]= &-& t \sum_{j~ n.n. i}[(1-n_{i,-s})\tilde c_{js} +
S_i^{\mp} \tilde c_{j,-s}]  + \nonumber \\
 &+&{1\over 2} J \sum_{j~ n.n. i}  (s S_j^z \tilde c_{i s} +
S_j^{\mp} \tilde c_{i,-s} ),  \quad s=\pm 1. \label{eq11}
\end{eqnarray}
It is convenient to express $n_{i,-s}=n_i/2 - sS_i^z$ in terms of
local density and spin operators. We are interested in the normal
metallic phase without any spin or charge long range order,
i.e. $\langle {\bf S}_i \rangle =0$ and $\langle n_i \rangle = c_e$.
Introducing $n_i =c_e + \tilde n_i$, we can rewrite Eq.(\ref{eq11}) in
the $\bf k$-representation,
\begin{eqnarray}
&&[\tilde c_{{\bf k} s},H]= -(1-{c_e\over 2}) t \gamma_{\bf k}\tilde 
c_{{\bf k} s} +
{t\over 2} \sum_{{\bf k}'} \gamma_{{\bf k}'} \tilde n_{{\bf k}-{\bf k}'}
\tilde c_{{\bf k}', s} + \nonumber \\
&+&\sum_{{\bf k}'} \bigl ( {J\over 2 } \gamma_{{\bf k}-{\bf k}'}  - t 
\gamma_{{\bf k}'}\bigr ) \bigl[ s S^z_{{\bf k}-{\bf k}'} 
\tilde c_{{\bf k}',s} + S^{\mp}_{{\bf k}-{\bf k}'} \tilde c_{{\bf k}',-s} 
\bigr], \label{eq12} 
\end{eqnarray}
where $\gamma_{\bf k}= \sum_{j~ n.n. 0} {\rm exp}(i{\bf k} \cdot {\bf
r}_j)$.

Using Eqs.(\ref{eq6},\ref{eq9},\ref{eq12}) we can express the 
`free'-propagation term  $\zeta_{\bf k}$ as
\begin{eqnarray}
\zeta_{\bf k}&=& {1\over \alpha} \langle \{[\tilde c_{{\bf k} s},H],
\tilde c^{\dagger}_{{\bf k} s}\}_+\rangle= \bar {\epsilon} +
\epsilon_{\bf k} ,\nonumber \\
\epsilon_{\bf k}&=&-{t\over \alpha} \sum_{j~n.n.i} \langle
(1-n_{i,-s})(1-n_{j,-s})\rangle {\mathrm e}^{i{\bf k} \cdot 
({\bf r}_i - {\bf r}_j)}, \label{eq13} \\
&=& \eta t \gamma_{\bf k}, \qquad \eta= \alpha + {1\over \alpha} 
\langle S_i^z S_j^z \rangle ,\nonumber
\end{eqnarray}
where  $\eta < \alpha$, due to AFM correlations. 

Note that the spectral function with $\alpha,\zeta_{\bf k}$,
following from Eqs.(\ref{eq9},\ref{eq10},\ref{eq13}), has correct 
lowest $l=0,1$ frequency moments
\begin{equation}
m_l =\int \omega^l A({\bf k},\omega) d\omega, \label{eq13a}
\end{equation}
while moments $m_{l>1}$ will not be exact due to approximations involved
in the calculation of $\Sigma({\bf k},\omega)$, described in the next
section.
 
\section{Self energy}

Eq.(\ref{eq12}) is determining the operator $C_{\bf k}$ corresponding
to Eq.(\ref{eq6}) and is the starting point for approximations to
$\Sigma({\bf k},z)$, Eq.(\ref{eq8}). The first term in Eq.(\ref{eq12})
is the `free' fermion propagation, which is absorbed into $\zeta_{\bf
k}$, Eq.(\ref{eq13}). The second and the third term represent coupling
of the fermion to density and spin fluctuations, respectively.  We are
interested mainly in the processes which at low $T\ll J,t$ dominate
the low-energy QP relaxation (damping). We still expect that at $T=0$,
or at least at low $T$ (assuming the metallic phase extending to low
$T$) there exists a well defined Fermi energy and the corresponding FS
determined by $\Sigma''({\bf k},\omega=0)=0$. The relaxation rate
$|\Sigma''({\bf k},\omega \ne 0)| >0$ is then dominated by the
coupling to low-energy spin and charge fluctuation modes.

It is well documented that the spin dynamics in the normal phase of
cuprates at $T>T_c$ is anomalous \cite{imai,shir}, with an enhanced
response at low $\omega$. This is best seen in the NMR relaxation on
$Cu$ sites with $T_1^{-1} \sim const$, \cite{imai} rather than
Korringa law $T_1^{-1}
\propto T$. So it seems evident that down to low $T \sim T_c$ 
spin fluctuations behave as quite independent degrees of freedom,
although not as a coherent mode. This naturally leads to the idea of
decoupling of spin dynamics from the fermion propagation. The density
fluctuations $\tilde n_{\bf k}$, on the other hand, seem to be less
important. One reason is that at low doping their scaling is with
$c_h$. Also there is so far no indication for pronounced low-$\omega$
density fluctuations, and the dynamics is restricted predominantly to
$\omega>t$ \cite{khal}.

The assumption that spin degrees behave as independent modes, leads to
the decoupling of correlations, e.g.,
\begin{eqnarray}
&&\langle S^z_{{\bf k}-{\bf k}'}(t)\tilde c_{{\bf k'}s}(t)
S^z_{{\bf k}''-{\bf k}}\tilde c^{\dagger}_{{\bf k}''s} \rangle \sim
\nonumber \\ 
&\sim&\delta_{{\bf k}' {\bf k}''}
\langle S^z_{{\bf k}-{\bf k}'}(t) S^z_{{\bf k}'-{\bf k}} \rangle
\langle \tilde c_{{\bf k}''s}(t)\tilde c^{\dagger}_{{\bf k}'s} \rangle,
\label{eq14}
\end{eqnarray}
and to the expression for  spin fluctuation contribution 
$\Sigma_{sf}({\bf k},\omega) =\langle\!\langle C_{\bf k};
C_{\bf k}\rangle\!\rangle_z^{irr}/\alpha $,
\begin{eqnarray}
\Sigma_{sf}({\bf k},z) =&&\sum_{{\bf k}'} M_{{\bf k k}'}
\int \int {d\omega_1 d\omega_2 \over \pi^2} g(\omega_1,\omega_2)
\times \nonumber \\
&&{A({{\bf k}'},\omega_1) \chi''({\bf k}-{\bf k}',\omega_2) \over
z-\omega_1-\omega_2 }, \label{eq15}
\end{eqnarray}
where $\chi$ is the dynamical spin susceptibility
\begin{equation}
\chi({\bf q},z)=-i\int_0^{\infty} {\mathrm e}^{izt} \langle
[S^z_{\bf q}(t) , S^z_{-\bf q}] \rangle, \label{eq16}
\end{equation}
and
\begin{eqnarray} 
M_{{\bf k} {\bf k}'} &=& {3 \over \alpha} (t \gamma _{{\bf k}'}
-{J \over 2} \gamma _{{\bf k}-{\bf k}'} )^2 , \label{eq17} \\
g(\omega_1,\omega_2)&=& 1- f(\omega_1) + \bar n(\omega_2),
\label{eq18}
\end{eqnarray}
where $f$ and $\bar n$ are Fermi and Bose distribution functions,
at $T=0$ leading to
\begin{eqnarray}
g(\omega_1,\omega_2)&=& 1,\quad \omega_1,\omega_2 >0, \nonumber\\
&=& -1,\quad \omega_1,\omega_2 <0, \label{eq19} \\
&=& 0 \quad \mathrm {otherwise}.\nonumber
\end{eqnarray}
In the derivation of Eq.(\ref{eq15}) we have assumed that in the spin
system there is no long range order or spin anisotropy,
i.e. $\chi^{\alpha\beta}({\bf k},z) =\delta_{\alpha\beta} \chi({\bf
k},z)$, and $G({\bf k},z)$ is spin invariant as well.

Let us briefly comment Eq.(\ref{eq15}). Similar expressions, coupling
QP dynamics to spin fluctuations, have been obtained for the Hubbard
model within the perturbation theory, or beyond that using
a diagrammatic approach \cite{kamp,bick}, as well as within the strongly
correlated $t-J$ model for the hole dynamics in the low-doping regime
\cite{plak}. Using a phenomenological starting point, analogous 
expressions for $\Sigma({\bf k},\omega)$ have been considered also
within the MFL theory \cite{litt} and in the nearly AFM scenario
\cite{mill}. It is important to note that in our theory the coupling 
$M$ involves besides $J$ explicitly also $t$, coming directly from
Eq.(\ref{eq11}), i.e. from the restriction of no double occupancy.

Eq.(\ref{eq15}) is intended primarily to describe the QP damping close
to the Fermi energy, i.e. $\Sigma''({\bf k},\omega \sim 0)$. On the
other hand, it is easy to establish, e.g. by calculating some higher
frequency moments $m_{l>1}$, Eq.(\ref{eq13a}), that $\Sigma_{sf}$ needs
corrections, in particular for $\omega \ll 0$. These arise due to
neglected density fluctuations in Eq.(\ref{eq12}), and also due to the
oversimplified decoupling, Eq.(\ref{eq15}). We remedy this deficiency
by adding a $\bf k$-independent contribution,
\begin{equation}
\Sigma({\bf k},z)= \Sigma_{sf}({\bf k},z)+\Sigma_{inc}(z). \label{eq20}
\end{equation}
The incoherent $\Sigma_{inc}$ has been studied in detail in connection
with a single hole introduced into the $J=0$ magnetic insulator
\cite{brin}. It persists also at $J>0$ \cite{schm} and at finite
doping \cite{dago,jakl2}. We will lateron choose
$\Sigma_{inc}(\omega)$ qualitatively consistent with previous studies,
introducing however two essential requirements on its form. First,
$\Sigma_{inc}$ should not influence the QP damping at $\omega \sim 0$
and not spoil the existence of the FS, i.e. $\Sigma''_{inc}(\omega \to
0) \propto \omega^{\nu}, ~\nu \agt 2 $. On the other hand, we require also
the conservation of the FS volume (Luttinger theorem)
within the correlated system \cite{lutt}. The latter seems to be
confirmed with ARPES experiments on cuprates \cite{shen} and also with
numerical studies of small model systems
\cite{step,dago,jakl2}. Note that the Fermi surface is determined by
conditions \cite{lutt}
\begin{equation}
\zeta_{{\bf k}_F} + \Sigma'({{\bf k}_F},0)= \mu, 
\qquad c_e = V_{FS}/V_0, \label{eq21}
\end{equation}
whereby the FS volume $V_{FS}$ should correspond to the fermion
density, where $V_0$ is the volume of the first Brilloiuin zone.  In
our case $c_e$ is given by
\begin{equation}
c_e={1\over N} \sum_{\bf k} \int_{-\infty}^0 A({\bf k}, \omega)d\omega.
\label{eq22}
\end{equation}
Working with fixed $c_e$ we can satisfy Eq.(\ref{eq22}) by choosing
appropriate $\mu$. Since our decoupling approximation is not
conserving automatically the FS volume, Eq.(\ref{eq21}), we can achieve
this only via certain restrictions on $\Sigma_{inc}$. We are mainly
interested in the regime $c_e
\alt 1$, which requires generally quite large $\Sigma_{inc}(\omega <0)$.

In the present work we do not intend to calculate spin
susceptibilities $\chi({\bf q},\omega)$, which enter
Eq.(\ref{eq15}). Rather we consider them as input. We assume
$\chi''({\bf q},\omega)$ according to recent numerical analysis of the
$t-J$ model \cite{jakl3} and consistent with anomalous spin
dynamics found in NMR \cite{imai} and neutron scattering experiments
on cuprates
\cite{shir}. Within the intermediate doping regime results
seem to imply the MFL form \cite{varm}, at least for the local
susceptibility at $T>0$, \cite{jakl3}
\begin{equation}
\chi_L''(\omega) = {1\over N} \sum_{\bf q} \chi''({\bf q},\omega)=
\mathrm {tanh} \bigl({\omega \over 2T}\bigr) 
~\bar S_L (\omega),
\label{eq23}
\end{equation}
where $\bar S_L(\omega)$ is essentially $T$-independent, and $\bar
S_L(\omega \to 0) =S_L^0 >0$. Since we consider in this paper only
$T=0$, this implies a nonanalytical $\chi''_L(\omega \sim 0) = {\rm
sign}(\omega) S_L^0$. At the same time we assume that the ${\bf
k}$-dependence of $\chi({\bf k},\omega)$ is less critical on
approaching $T=0$. The latter assumption distinguishes the MFL theory
\cite{varm,litt} from some other scenarios\cite{mill}.

After choosing particular forms for $\chi({\bf q},\omega)$ and
$\Sigma_{inc}(\omega)$, and at fixed model parameters and $c_h$, we
are faced with a closed set of selfconsistent (SC) equations
(\ref{eq9},\ref{eq15}), which we discuss in the following section.

\section{Results} 

We apply our analysis to the case of the $t-J$ model on 2D square
lattice, as relevant for strongly correlated electrons in doped AFM,
and particularly in cuprates \cite{rice}.  This implies the regime
$J<t$ and $c_e \alt 1$. Due to specific approximations our approach is
applicable in a restricted window of $c_h$. We expect that this regime
corresponds to the intermediate doping. For very low doping $c_h
\ll 1$ electrons are closer to a description in terms of
independent AFM spin polarons \cite{schm}, where it is hard to
establish the existence of the FS \cite{jakl2}, and the spin
background is nearly an ordered AFM state. Such facts are not properly
incorporated within our decoupling approximation. On the other side,
within the strongly `overdoped' regime the spin fluctuations loose
their identity as independent degrees of freedom. Within the $t-J$
model the `optimum' doping is plausibly determined by the competition
between the kinetic energy density $\propto c_h t$ and the AFM
exchange energy density $\propto J$, so that the `intermediate' regime
is for us a broad range around $c_h \sim J/t$.  We furtheron
consider only $T=0$.

Since we are interested only in a qualitative description of QP, we
assume the simplest form for dynamical spin susceptibility $\chi({\bf
q},\omega)$, consistent with the MFL form for the $\omega$-dependence,
Eq.(\ref{eq23}), while the $\bf q$-dependence is determined by the AFM
inverse correlation length $\kappa$,
\begin{equation}
\chi''({\bf q},\omega)={W \over (|{\bf q}-{\bf Q}|^2 +\kappa^2)
(\omega^2+ \omega^2_0)} \mathrm {sign}(\omega), \label{eq40}
\end{equation}
where ${\bf Q}=(\pi/2,\pi/2)$ is the AFM wavevector, and
the characteristic exchange frequency $\omega_0 \propto J$. 
$W$ is estimated via the sum rule
\begin{eqnarray}
{1\over N} \sum_{\bf q} \int_0^{\infty} \chi''({\bf q},\omega)
d \omega = \langle S_i^z\rangle^2 = {1\over 4} (1-c_h),
\label{eq41}
\end{eqnarray}
For $\kappa <1$ this yields $W\sim \omega_0 \kappa^2/2$. A similar
form to Eq.(\ref{eq40}) has been indeed found in the numerical 
analysis of finite systems \cite{jakl2}.

Let us consider some qualitative consequences of the theory.  First we
discuss the hole part $\omega<0$, corresponding to the photoemission
(PES) spectra \cite{shen}.  The crucial contribution to
$\Sigma''_{sf}({\bf k},\omega \sim 0)$ in Eq.(\ref{eq15}) comes from
spin fluctuations with ${\bf q} \sim {\bf Q}$ and ${\bf k},{\bf k}'
\sim {\bf k_F}$. Hence we can estimate
\begin{equation}
M_{{\bf k k}'} \sim M_0 = {3\over \alpha} (t \gamma_{{\bf k}_F}
+ 2J)^2,\qquad  \gamma_{{\bf k}_F} \sim {4c_h \over \lambda},
\label {eq42}
\end{equation}
and $\lambda \agt 1$. Since on a square lattice at $c_h \ll
1$ the FS is nearly nested with a wavevector $\tilde {\bf Q}$,
$\tilde Q \alt Q$, we get from Eqs.(\ref{eq15},\ref{eq40}) for
$|\omega| < \omega_0$
\begin{eqnarray}
\Sigma''({\bf k}_F,\omega) &\sim& -M_0 \mathrm 
\chi''(\tilde {\bf Q},\omega)  \int^{\omega}_0
{d\omega_1 \over \pi} A({\bf k}_F-\tilde {\bf Q},\omega_1) \nonumber
\\ &\propto& -{M_0 {\cal N}_F\over \omega_0} |\omega| ,\label{eq43}
\end{eqnarray}
provided that $\tilde Q-Q < \kappa$, and ${\cal N}_F$ is the density
of states at the FS.  It is not surprising that we recover the MFL
form $\Sigma''(\omega) \propto |\omega|$, since Eq.(\ref{eq15})
has a close analogy to the phenomenological derivation
\cite{litt}, where the $\omega$-dependence of the boson
spectrum was assumed the same as in Eq.(\ref{eq40}). The essential
difference is that within the phenomenological theory {\bf
k}-dependences are neglected. In our approach the $\bf q$-dependence
of spin fluctuations is important, and we expect the MFL form only if
the approximate nesting condition is fulfilled, taking into account
the width of the maximum $\sim \kappa$ in Eq.(\ref{eq40}).

When we estimate $r=\Sigma''({\bf k}_F,\omega)/\omega$ from
Eq.(\ref{eq43}) for the `optimum' doping $c_h \sim J/t$, it is
characteristic that $|r| >1$. This means that the hole-like QP peaks
are generally overdamped (with QP damping larger than the
characteristic QP frequency), as indeed observed in recent numerical
studies \cite{jakl2}, and consistent with ARPES experiments on
cuprates \cite{shen}.

Another consequence of Eq.(\ref{eq15}) is the asymmetry between the
hole part, $\omega<0$, and the electron part $\omega>0$ of the
spectra, also evident in numerical studies \cite{jakl2}. Within the
present theory this phenomenon originates partially from $M_{{\bf k
k}'}$, Eq.(\ref{eq17}), which reaches minimum effectively for
$k>k_F$. Unlike in $M_0$, Eq.(\ref{eq42}), where both terms inside the
bracket have the same sign and are of the same order of magnitude,
terms in Eq.(\ref{eq17}) are cancelling each other for
$k>k_F$. Another difference comes from $\Sigma''_{inc}(\omega)$ which
should be added predominantly at $\omega<0$, in order to satisfy the
conservation of $V_{FS}$, Eq.(\ref{eq21}). This enhances strongly only
the QP damping within the hole part.
 
Let us finally turn to the numerical analysis of SC equations.
Here we present only some general features of QP
spectra $A({\bf k},\omega)$.
To be close to the regime of cuprates we choose $J = 0.3~t$
\cite{rice} and $\omega_0 \sim 2J$ \cite{jakl3}, while
the lattice is a 2D square one. Although it is necessary to
acknowledge the importance of short range AFM correlations even in the
regime of intermediate doping, we assume for simplicity in
Eq.(\ref{eq13}) $\eta=\alpha$. Namely, in Eq.(\ref{eq13}) AFM
correlations enter only to modify the width of the effective band
$\epsilon_{\bf k}$, and thus have a less crucial quantitative
effect. We also put $\bar \epsilon=0$. We are interested in the
intermediate doping regime with short range AFM correlations
$\xi=\kappa^{-1} \sim 1$.  For fixed $\mu$ (which has some technical
advantage over fixing $c_h$) the numerical task is now to calculate
$c_h$ from Eq.(\ref{eq22}) and to use such $\Sigma_{inc}$ to
simultaneously satisfy Eq.(\ref{eq21}). The form of
$\Sigma''_{inc}(\omega)$ in our theory is quite arbitrary provided
that $\Sigma_{inc}''(0)=0$, since it enters Eq.(\ref{eq21}) only via
$\Sigma'_{inc}(0)$. In the calculation we choose a simple parabolic
form for $\Sigma''_{inc}(\omega)$ within the interval
$[\omega_1,\omega_2],~ \omega_1<\omega_2<0$ with the maximum value
max$|\Sigma''_{inc}(\omega)|=\Sigma_0$ as a parameter. To eliminate
discontinuities we smoothly connect the latter parabola with another
one in the interval $[\omega_2,0]$ following the analytic FL form
$\Sigma''_{inc}(\omega)=-c\omega^2$.

In the following we present the specific case with $c_h=0.3$, where
we fix $\xi=1.0$. Choosing
$\omega_1/t=-6,~\omega_2/t=-1$ we still have to use quite large
$\Sigma_0=3.5~t$ to satisfy Eq.(\ref{eq22}). In Fig.1 we present 
$A({\bf k},\omega)$ for selected direction ${\bf k}=x (\pi,\pi),~~0<x<1$, 
and in Figs.2,3 the corresponding $\Sigma''({\bf k},\omega)$, and $\tilde
\Sigma'({\bf k},\omega)=\Sigma'({\bf
k},\omega) +\zeta_{\bf k}-\omega-\mu $.  We can comment on several
characteristic features.  While the damping $\Sigma''({\bf
k},\omega\ll 0)$ is dominated by the incoherent contribution
$\Sigma''_{inc}$, the behaviour at $\omega \sim 0$ is governed
by the spin-fluctuation part $\Sigma''_{sf}$. It is evident that for
${\bf k}
\sim {\bf k}_F$ (here the FS is at $x \sim 0.42$) the form is MFL-like
$\Sigma''({\bf k},\omega<0) \sim - r_{\bf k} |\omega|$, where $r_{\bf
k}$ is increasing with $k$. On the other hand, the damping becomes
more FL like $\Sigma''({\bf k},\omega)\propto
\omega^2$ for $k \ll k_F$ as well as for $k \gg k_F$.
Both these facts and the observation that $r_{\bf k}>1$ are consistent
with recent numerical results within the $t-J$ model
\cite{jakl2}. Note that also the magnitude and the overall shape of
$\Sigma''({\bf k},\omega)$ is in agreement with numerical results
\cite{jakl2}, supporting in addition our choice of $\Sigma''_{inc}$.
Evident from Fig.1 is also the prononunced asymmetry between the hole
part $\omega<0$ and the electron part $\omega>0$. QP excitations are
generally overdamped for $k<k_F$, and moreover immersed in an
incoherent background for $k\ll k_F$. On the other hand, the QP peaks
for $k>k_F$ are better prononunced and less damped
\cite{jakl3}.  It should be however noted that for $\omega>0$ the
maximum in the damping $|\Sigma''(k>k_F,\omega)| \sim t$, seen in
Fig.2, is probably overestimated within our decoupling. Hence in Fig.1
the QP peaks at $k\gg k_F$ seem to be more damped than found in
the numerical study \cite{jakl2}. 

In Fig.4 we show also the corresponding density of states ${\cal
N}(\omega) = (2/N)\sum_{\bf k}A({\bf k},\omega)$, which should be in
general less sensistive to strong correlations, and its bandwidth
close to the free fermion case.  ${\cal N}(\omega \ll 0)$ is
clearly dominated by the incoherent part $\Sigma_{inc}''$. A peculiar
maximum (which is not a singularity) appears at $\omega \agt 0$ and is
 related to the minimal damping just above FS.

The qualitative features remain similar for a whole range of
intermediate dopings $0.15<c_h<0.4$. It is however evident that on
approaching $c_h<0.2$ it becomes harder to satisfy the FS requirement,
Eq.(\ref{eq21}), and the results depends stronger on the form chosen
for $\Sigma_{inc}''$. This is plausible since $G({\bf k},\omega)$
becomes dominated by the incoherent contributions, studied extensively
in connection with a single mobile hole \cite{brin,schm}. In our
approximation the treatment of such an incoherent motion is too
crude. Nevertheless we can still study low $\omega \sim 0$ dynamics,
if we relax somewhat the condition Eq.(\ref{eq21}).

\section{Conclusions}

In this paper we have presented a simple theory for the electron
Green's function in the doped AFM. Note that our treatment differs
essentially from the analysis of a single hole in an AFM, or
slave-boson approaches for finite doping, where the spinless hole
(holon) propagators are studied in the first place.  We study directly
the equations of motion for electron operators.  This prevents the use
of usual diagrammatic techniques. Still we argue that in the sense of
mode-coupling theories our approximations, decoupling the self energy
into the single-electron propagation and spin fluctuations, are
reasonable at low $\omega$, as far as the spin fluctuations behave as
independent degrees of freedom.  It is well possible that such
assumption breaks down at certain energy or temperature scale
$\omega^* \sim T^*$.  Nevertheless such $\omega^* \ll t,J$ is clearly
quite low, so far not seen in small-system studies \cite{jakl2}, and
also hard to resolve from e.g. ARPES experiments
\cite{shen}, but it is possibly related to the onset of 
superconductivity in cuprates.

The main message of our theory, feasible for the intermediate-doping
regime, is that MFL-type spin dynamics implies also the MFL-type QP
damping and related spectral shapes.  The coupling to spin
fluctuations is necessarily strong, since it is related to both $J$
and $t$, Eq.(\ref{eq42}), and consequently QP features in the hole
part are generally overdamped.  Our analysis relies on the existence
of the well defined FS and on the conservation of its volume
(Luttinger theorem). It should be however noted, that the obtained
(large) FS within the $t-J$ model is close to a circular one, unlike
the experimental ones in cuprates \cite {shen}. It is probably easy to
change the shape of FS by including additional terms in the model
Eq.(\ref{eq1}), e.g. by introducing the n.n.n. hopping.
  
It is clear that the present theory is not fully selfconsistent, since
it does not describe the origin of the anomalous spin dynamics, which
seems to be the essential ingredient and also challenge within the
intermediate-doping regime, and could arise again from the frustration
induced by mobile holes.

\acknowledgements

The authors wishes to thank P. Horsch, G. Khaliullin and
R. Zeyher for useful suggestions and fruitful
discussions, and acknowledges the support of the MPI f\"ur
Festk\"orperforschung, Stuttgart, where a part of this work has been
performed.

\begin{figure}
\caption{ Spectral functions $A({\bf k},\omega)$ for $c_h=0.3$ and
$J/t=0.3$, for various ${\bf k}= x(\pi,\pi)$. $x$ are presented in 
steps of $0.1$. }
\label{fig1}
\end{figure}

\begin{figure}
\caption{ Imaginary part of the self energy $\Sigma''({\bf k},\omega)$,
corresponding to results on Fig.~1. }
\label{fig2}
\end{figure}

\begin{figure}
\caption{ Real part of the self energy $\tilde \Sigma'({\bf
k},\omega)$ =
$\Sigma'({\bf k},\omega) +\zeta_{\bf k} -\omega -\mu$, corresponding
to Fig.~1. } 
\label{fig3}
\end{figure}

\begin{figure}
\caption{ Density of states ${\cal N}(\omega)$, for data as above. }
\label{fig4}
\end{figure}
     
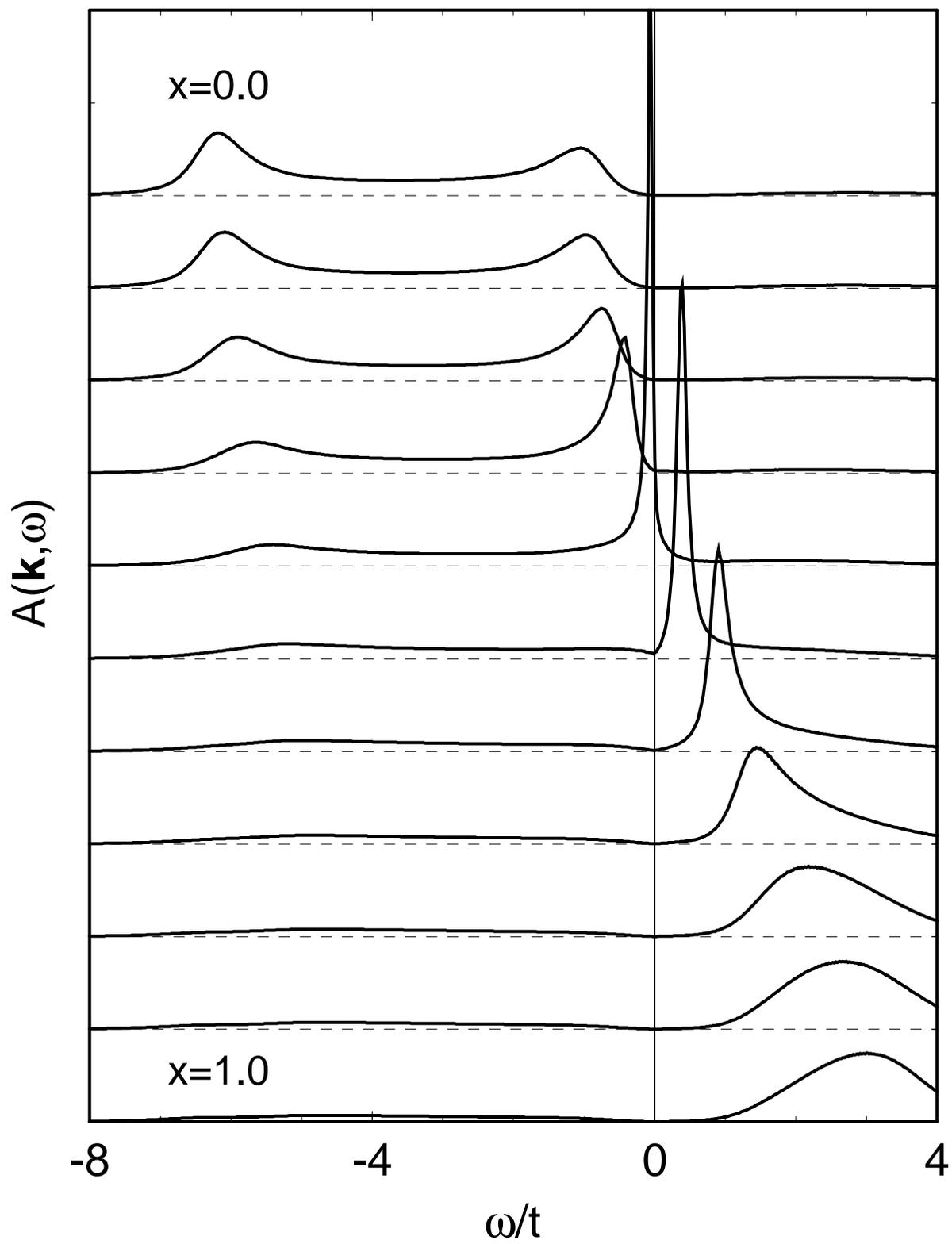
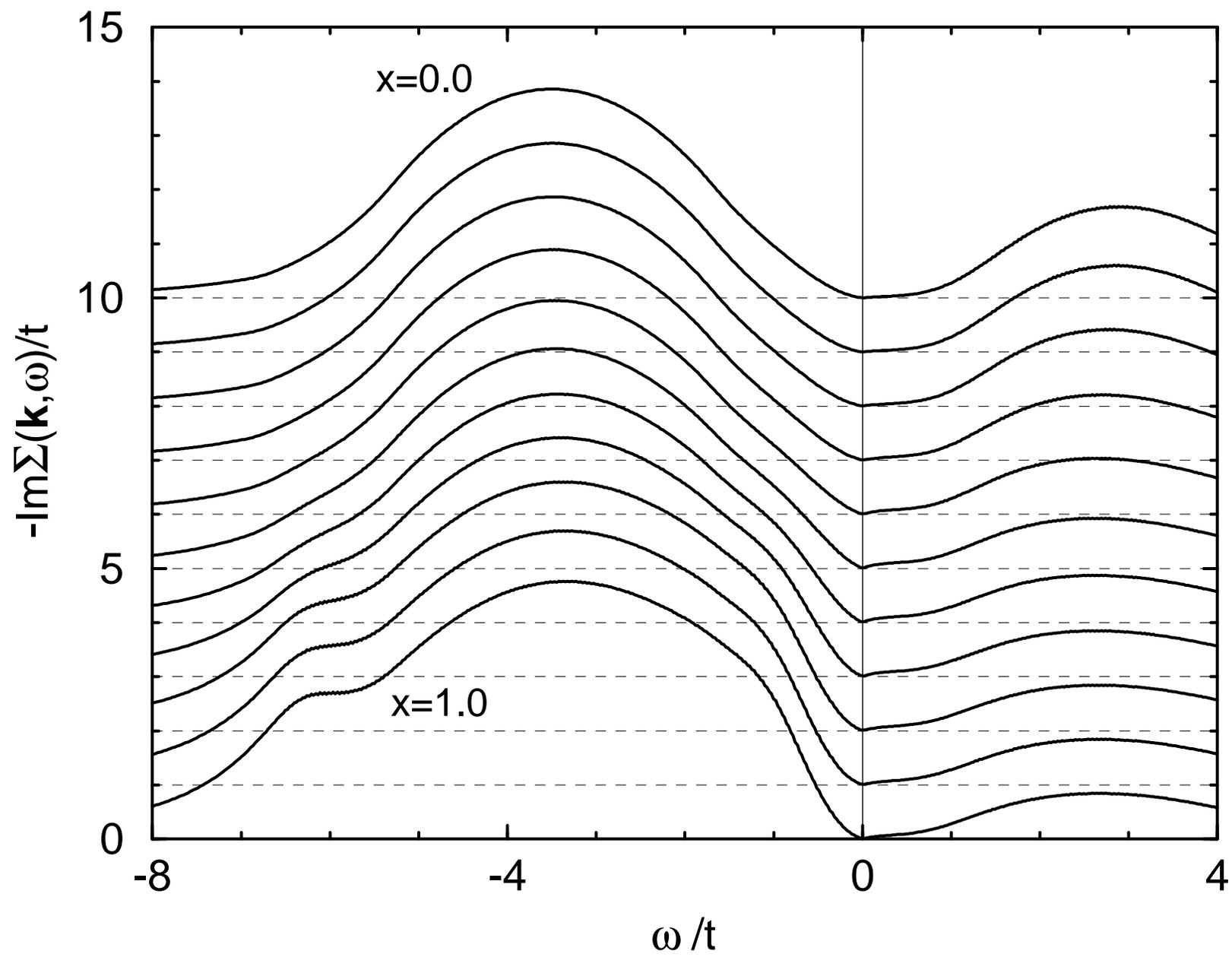
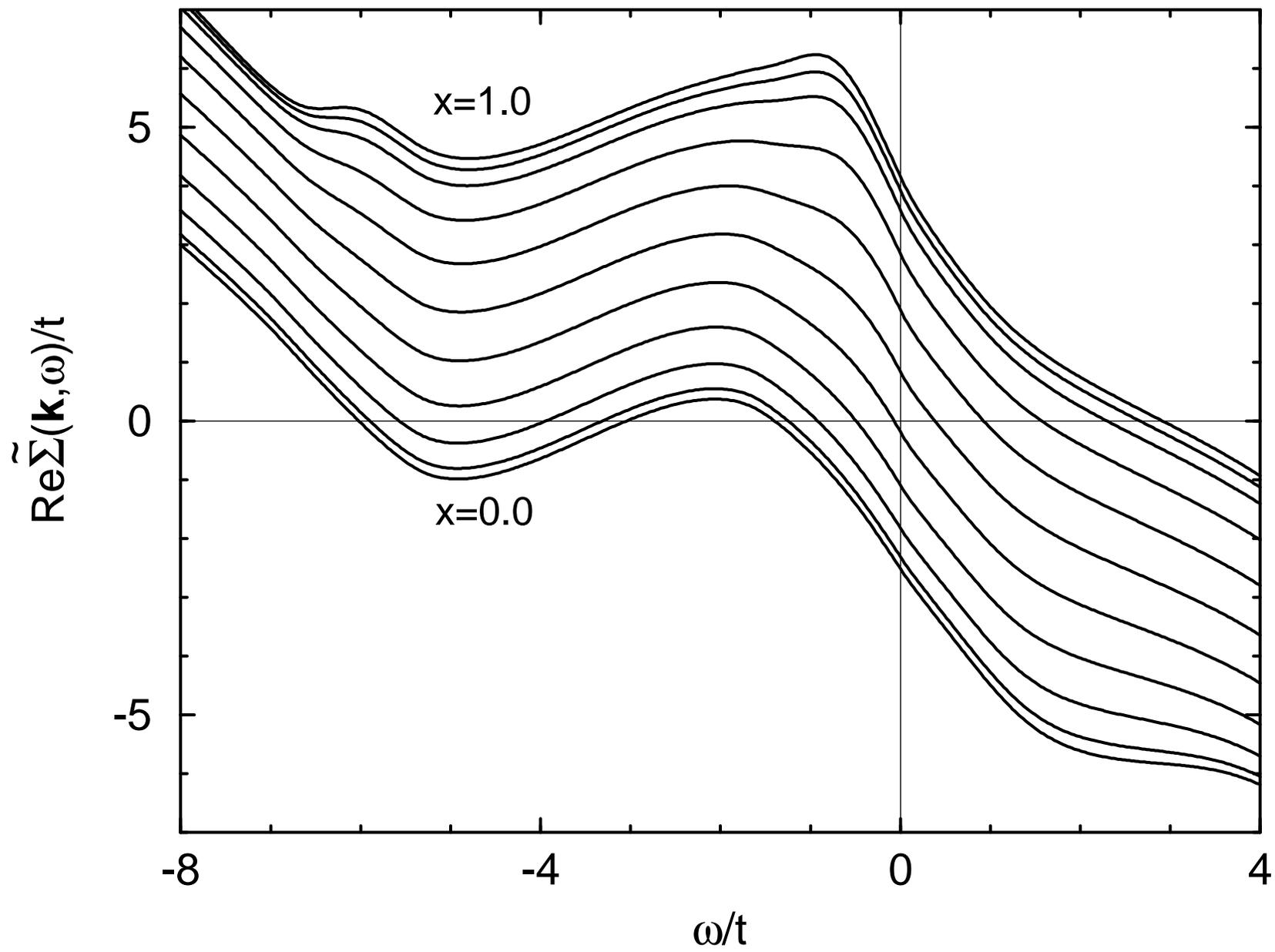
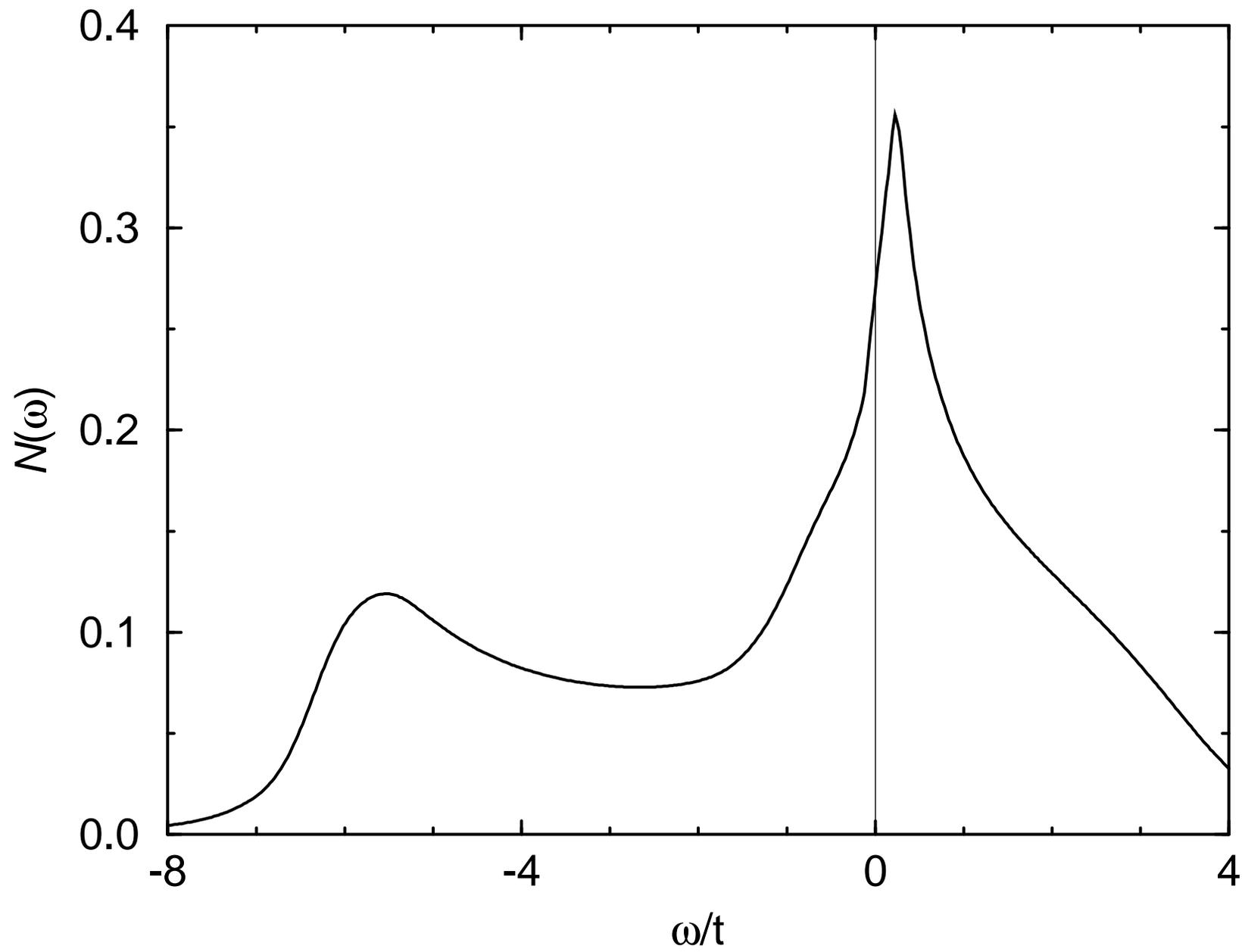
\end{document}